\documentclass[fleqn,twoside]{article}
\usepackage{espcrc2}

\usepackage{graphicx}
\usepackage[figuresright]{rotating}

\usepackage{amsmath,amssymb}

\title{Excited $B$ mesons from the lattice}

\author{J. Koponen\address[HY]{Department of Physical Sciences, \\ 
        P.O. Box 64, 00014 University of Helsinki, Finland}%
        \thanks{In collaboration with A. M. Green,
J. Ignatius, M. Jahma, C. McNeile, C. Michael and G. Thompson.
This work was supported by the Center for Scientific Computing,
Espoo, Finland, the Finnish Cultural Foundation, the Magnus Ehrnrooth
Foundation, the Academy of Finland (project 54038) and the EU
(grant HPRN-CT-2002-00311).}, for UKQCD Collaboration
}

\begin{document}

       
\begin{abstract}
The energies of different angular momentum states of a
heavy-light meson were measured on a lattice in \cite{MP}.
We have now repeated this study using several different
lattices, quenched and unquenched, that have different
physical lattice sizes, clover coefficients and quark--%
gluon couplings. The heavy quark is taken to be infinitely
heavy, whereas the light quark mass is approximately that 
of the strange quark. By interpolating and extrapolating
in the light quark mass we can thus compare the lattice
results with $B$ and $B_s$ meson experiments. Most interesting
is the lowest P-wave $B_s$ state, since it is possible
that it lies below the $BK$ threshold and hence is very
narrow. Unfortunately, there are no experimental results
on P-wave $B$ or $B_s$ mesons available at present.

In addition to the energy spectrum, we measured earlier also vector
(charge) and scalar (matter) radial distributions of the light quark
in the S-wave states of a heavy-light meson on a lattice \cite{GKMP}.
Now we are extending the study of radial distributions to P-wave states.
\vspace{1pc}
\end{abstract}

\maketitle

\section{Energies}

\begin{figure}
\centering
\caption{a) Two-point correlation function $C_2$;
b) Three-point correlation function $C_3$.}
\begin{tabular}{cc}
\includegraphics*[width=0.19\textwidth]{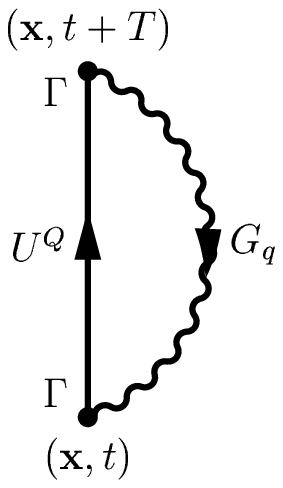}&
\includegraphics*[width=0.19\textwidth]{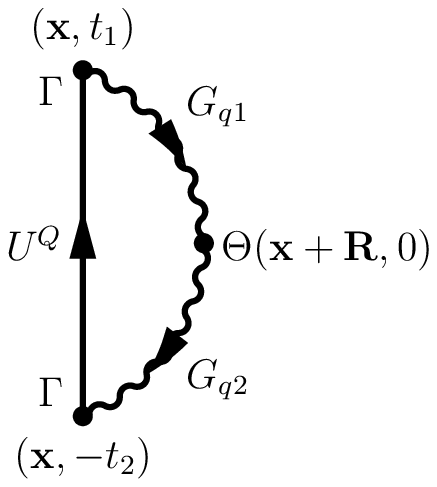}\\
 a) & b) \\ 
\end{tabular}
\label{fig_C2C3}
\end{figure}

The basic quantity for evaluating the energies of heavy-light mesons is
the 2-point correlation function --- see Fig.~\ref{fig_C2C3} a).
It is defined as
\begin{multline}
C_2(T)=\langle P_t\Gamma G_q(\mathbf{x},t+T,t)\\
\cdot P_{t+T}\Gamma^{\dag}U^Q(\mathbf{x},t,t+T)\rangle,
\end{multline}
where $U^Q(\mathbf{x},t,T)$ is the heavy (infinite mass)-quark propagator
and $G_q(\mathbf{x},t+T,t)$ the light anti-quark propagator. $P_t$
is a linear combination of products of gauge links at time $t$
along paths $P$ and $\Gamma$ defines the spin structure of the operator.
The $\langle ...\rangle$ means the average over the whole lattice.
The energies are then extracted by fitting the $C_2$
with a sum of exponentials,
\begin{equation}
C_2(T)\approx\sum_{i=1}^{N_{\textrm{max}}}c_{i}\mathrm{e}^{-m_i T},
\label{equ_C2}
\end{equation}
where $N_{\textrm{max}}=2\textrm{ -- }4$, $T\leq 10$.

\begin{table}
\centering
\caption{Lattice parameters.}
\begin{tabular}{cccc}
\hline
   &$\beta$&$C_{\textrm{SW}}$&$\kappa$\\
\hline
Q3 &5.7    &1.57             &0.14077 \\
DF2&5.2    &1.76             &0.1395  \\
DF3&5.2    &2.0171           &0.1350  \\
\hline
\hline
   &$a$ [fm]&$m_q/m_s$&$r_0m_\pi$\\
\hline
Q3 &0.179(9)    &0.83     &1.555(6)\\
DF2&0.152(8)    &1.28     &1.94(3)\\
DF3&0.110(6)    &1.12     &1.93(3)\\
\hline
\end{tabular}
\label{tbl_lattparam}
\end{table}

\begin{figure}
\caption{The energy spectrum for different angular momentum states.
 Here L$+$($-$) means that the light quark spin couples to angular
 momentum L giving the total $j=L\pm 1/2$. 2S is the first radially
 excited $L=0$ state. The spectrum was measured using different lattices:
 Q3 (a $16^3\times 24$ quenched lattice, \cite{MP}) and DF2, DF3
 ($16^3\times 24$ and $16^3\times 32$ lattices respectively, dynamical
 fermions, \cite{GKMMT}). Here $r_0=0.525(25)$~fm. Lattice parameters
 are given in Table~\ref{tbl_lattparam}. Note that the spectrum
 shows an approximately linear rise in excitation energy with L.}
\centering
\includegraphics*[angle=-90,width=0.46\textwidth]{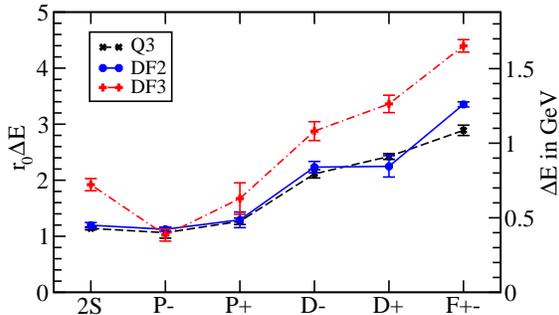}
\label{fig:energy}
\end{figure}

\begin{figure}
\caption{Interpolation from infinite mass to $m_Q=m_b$.
 The results at $m_Q=m_c$ are from experiments. Note that the
 lowest $B_s$ P-wave state lies below the $BK$
 threshold and should be very narrow.}
\centering
\includegraphics*[angle=-90,width=0.46\textwidth]{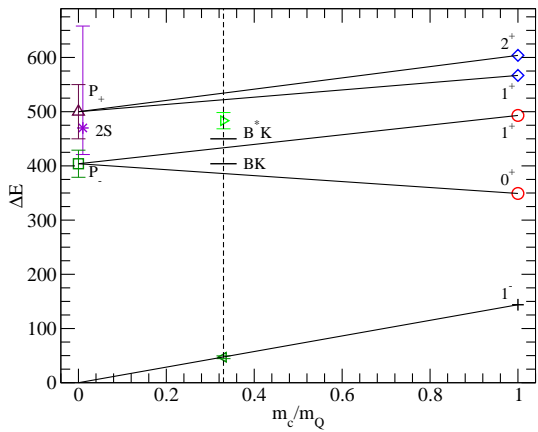}
\end{figure}

From some theoretical considerations \cite{Schnitzer}
it is expected that, for higher angular momentum states, the
multiplets should be inverted compared with the Coulomb spectrum,
\textit{i.e.} L$-$ should lie higher than L$+$ (see caption of
Fig.~\ref{fig:energy} for notation). Experimentally this inversion
is not seen for P-waves, and now the lattice measurements
show that there is no inversion in the D-wave states either. In fact,
the D$+$ and D$-$ states seem to be nearly degenerate, \textit{i.e.}
the spin-orbit splitting is very small (see Fig.~\ref{fig:energy}).

\section{Radial distributions}

\begin{figure}
\caption{Ground state vector and scalar radial distributions and discretized
 exponential fits.}
\vspace{-4mm}
\begin{center}
\includegraphics*[angle=-90,width=0.37\textwidth]{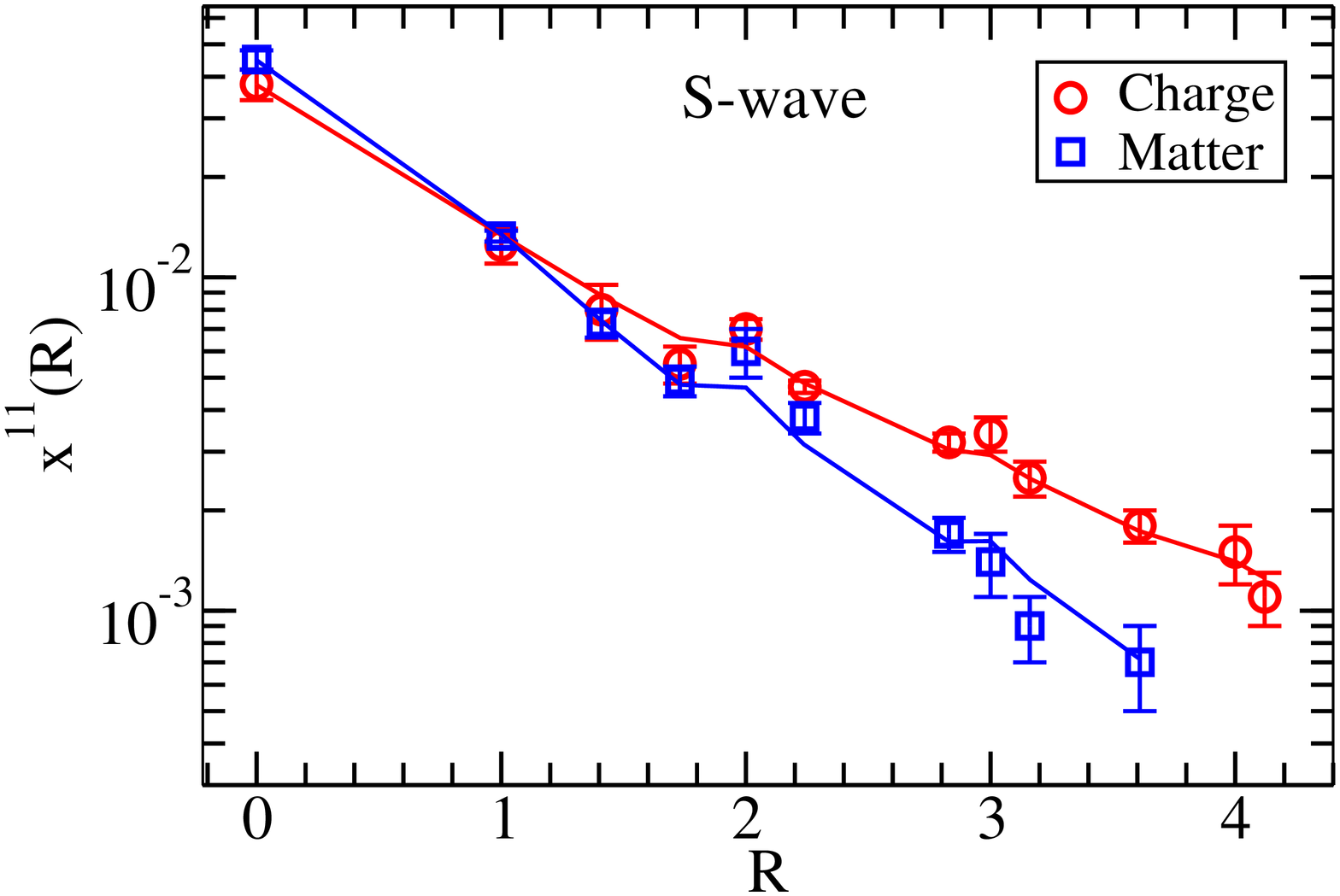}
\includegraphics*[angle=-90,width=0.37\textwidth]{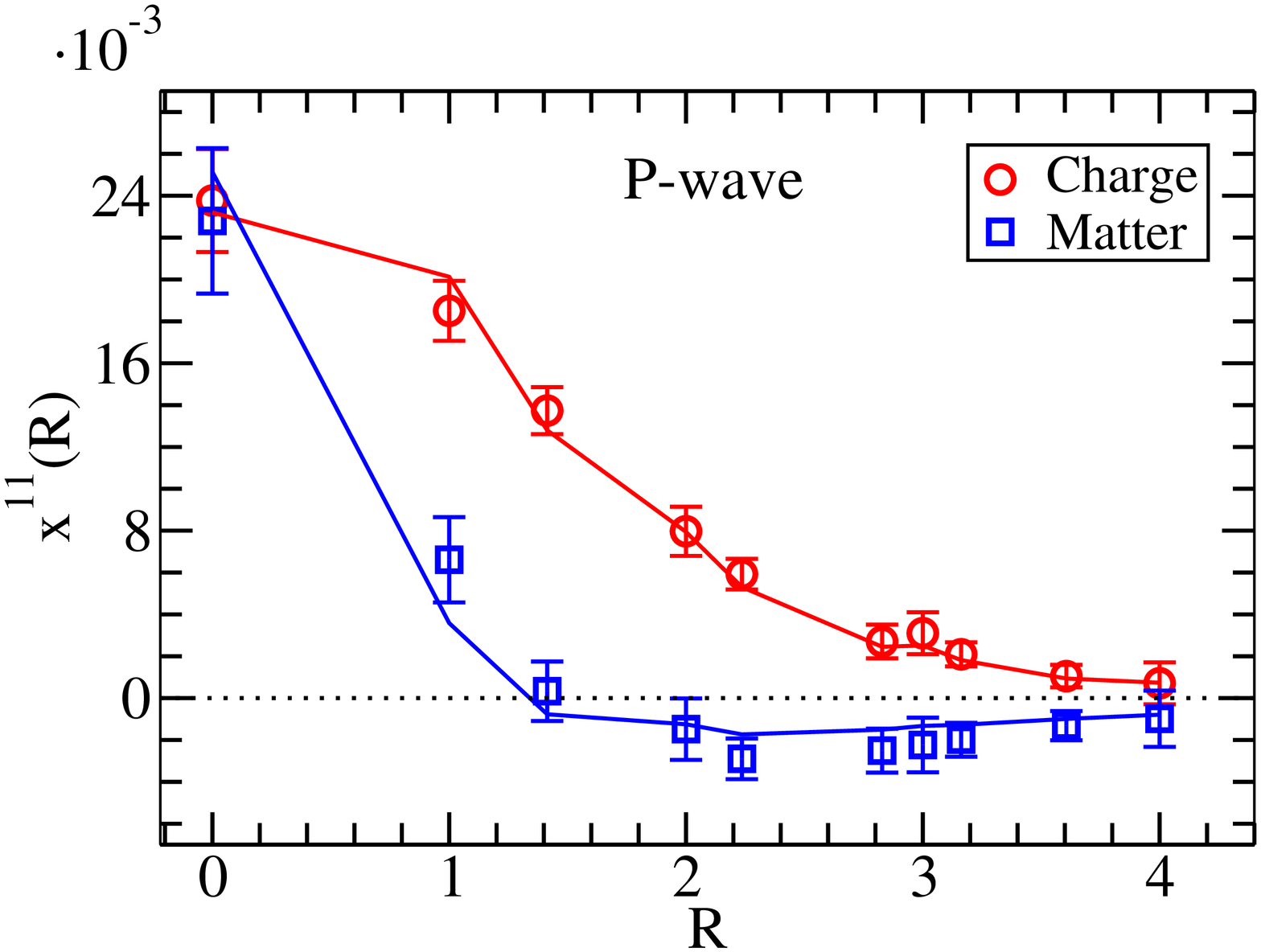}
\end{center}
\label{fig_X11}
\vspace{-8mm}
\end{figure}

\begin{figure}
\caption{Radial distributions containing the first radially
 excited state (Ref.~\cite{GKMP} for the S-wave).}
\vspace{-4mm}
\begin{center}
\includegraphics*[angle=-90,width=0.37\textwidth]{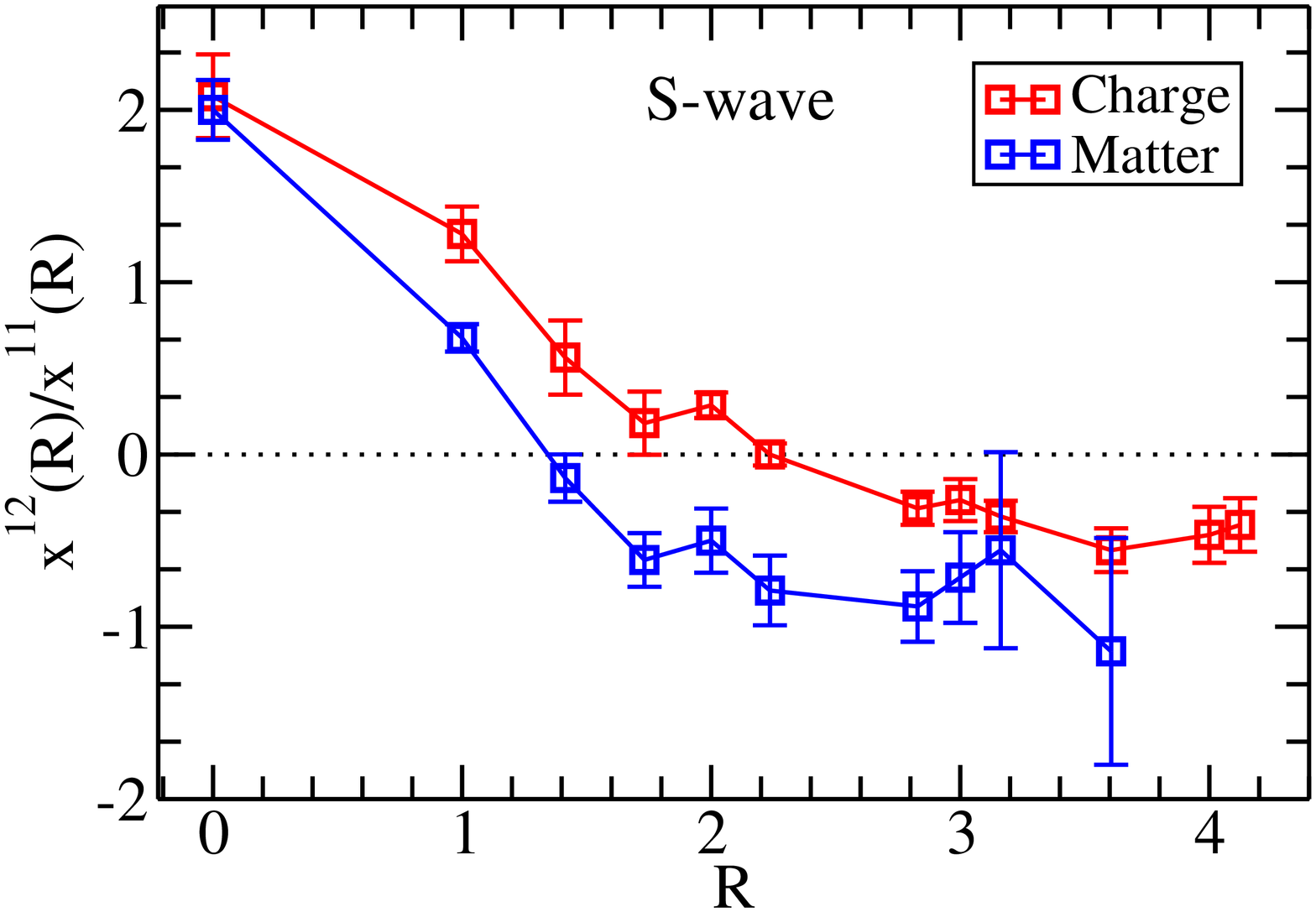}
\includegraphics*[angle=-90,width=0.37\textwidth]{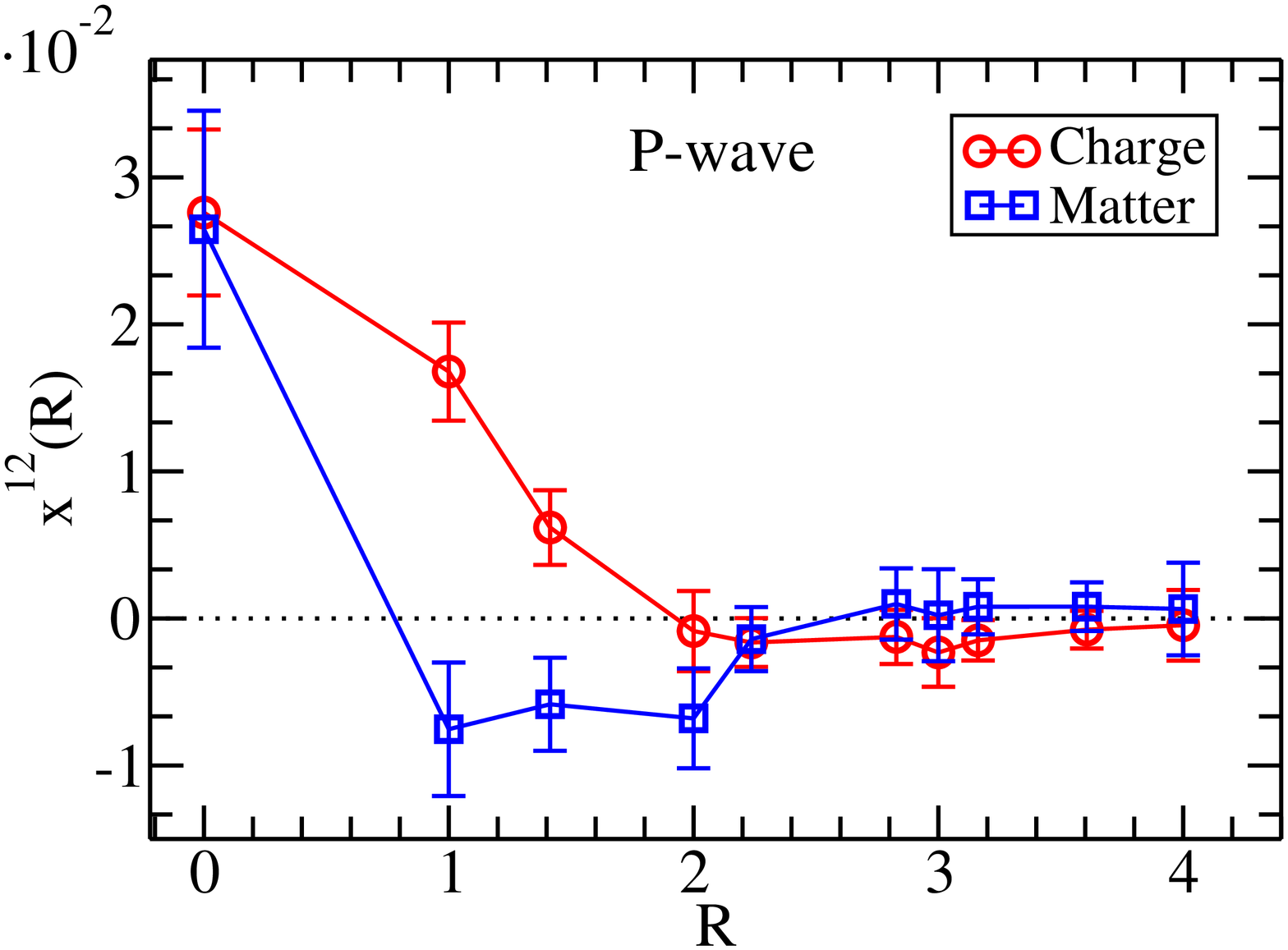}
\end{center}
\label{fig_X12}
\end{figure}

For evaluating the radial distributions of the light quark
a three-point correlation function is needed --- see
Fig.~\ref{fig_C2C3} b). It is defined as
\begin{equation}
C_3(R,T)=\langle \Gamma^{\dag}U^Q\Gamma G_{q1} \Theta
G_{q2}\rangle.
\end{equation}
We have now two light quark propagators, $G_{q1}$ and $G_{q2}$,
and a probe $\Theta(R)$ at distance $R$ from the
static quark. We have used two probes:
$\gamma_4$ for the vector (charge) and $1$
for the scalar (matter) distribution. The radial distributions,
$x^{ij}(R)$'s, are then extracted by fitting the $C_3$ with
\begin{equation}
C_3(R,T)\approx\sum_{i,j=1}^{N_{\textrm{max}}}c_{i}\mathrm{e}^{-m_i t_1}%
x^{ij}(R)\mathrm{e}^{-m_j t_2}c_{j}
\end{equation}
--- see Figs.~\ref{fig_X11}, \ref{fig_X12}. The $m_i$'s and $c_i$'s
are from Eq.~\ref{equ_C2}.

\begin{figure}
\caption{Dirac equation numerical solutions for a potential
 $V=-e/R+b\cdot R$ compared with lattice results. Here
 $e=0.6\cdot\hslash c$, $b=(500~\textrm{MeV})^2$ and $m=100$~MeV.
 $R$ is in lattice units ($a \approx 0.15$~fm).}
\vspace{-4mm}
\begin{center}
\includegraphics*[angle=-90,width=0.37\textwidth]{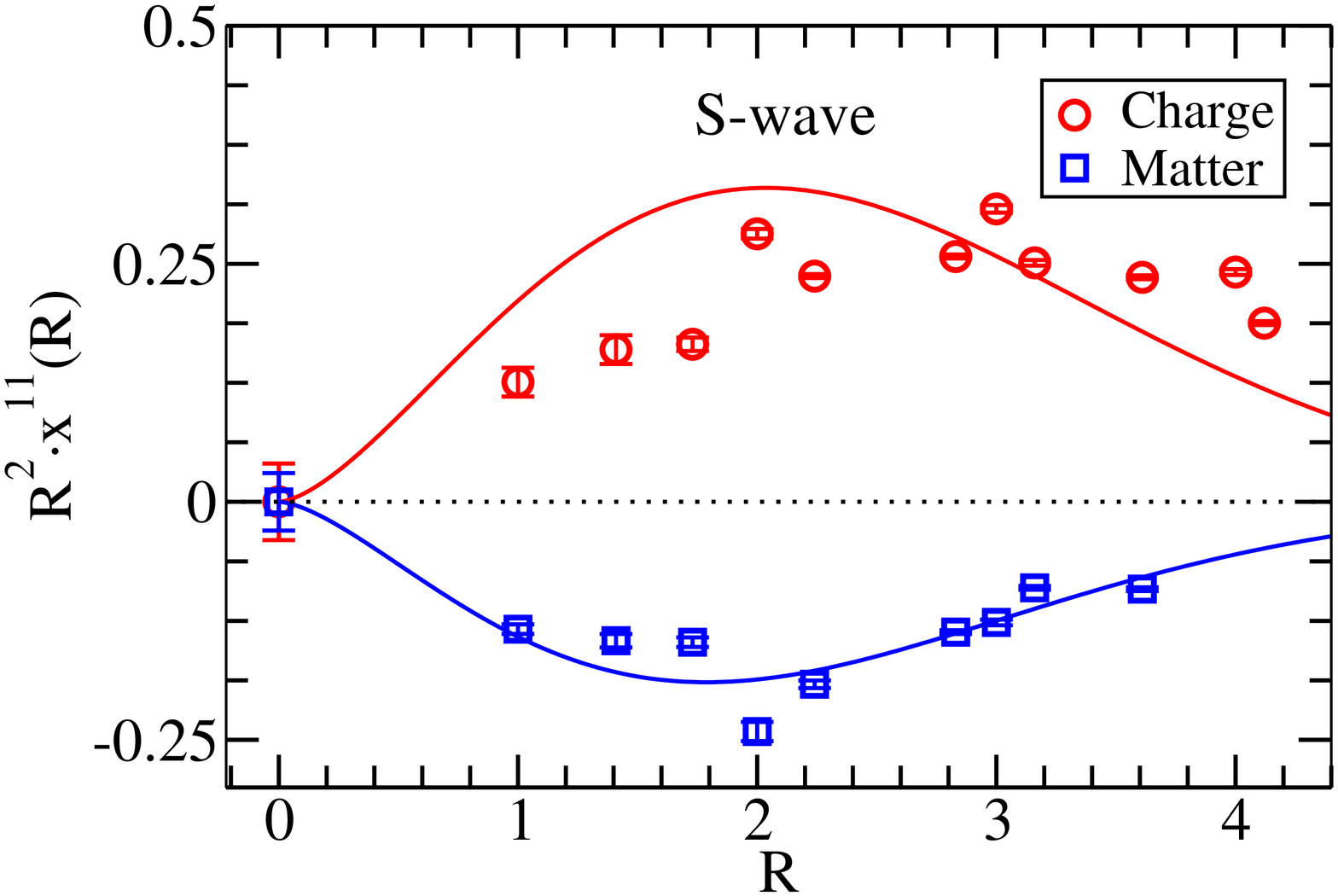}
\includegraphics*[angle=-90,width=0.37\textwidth]{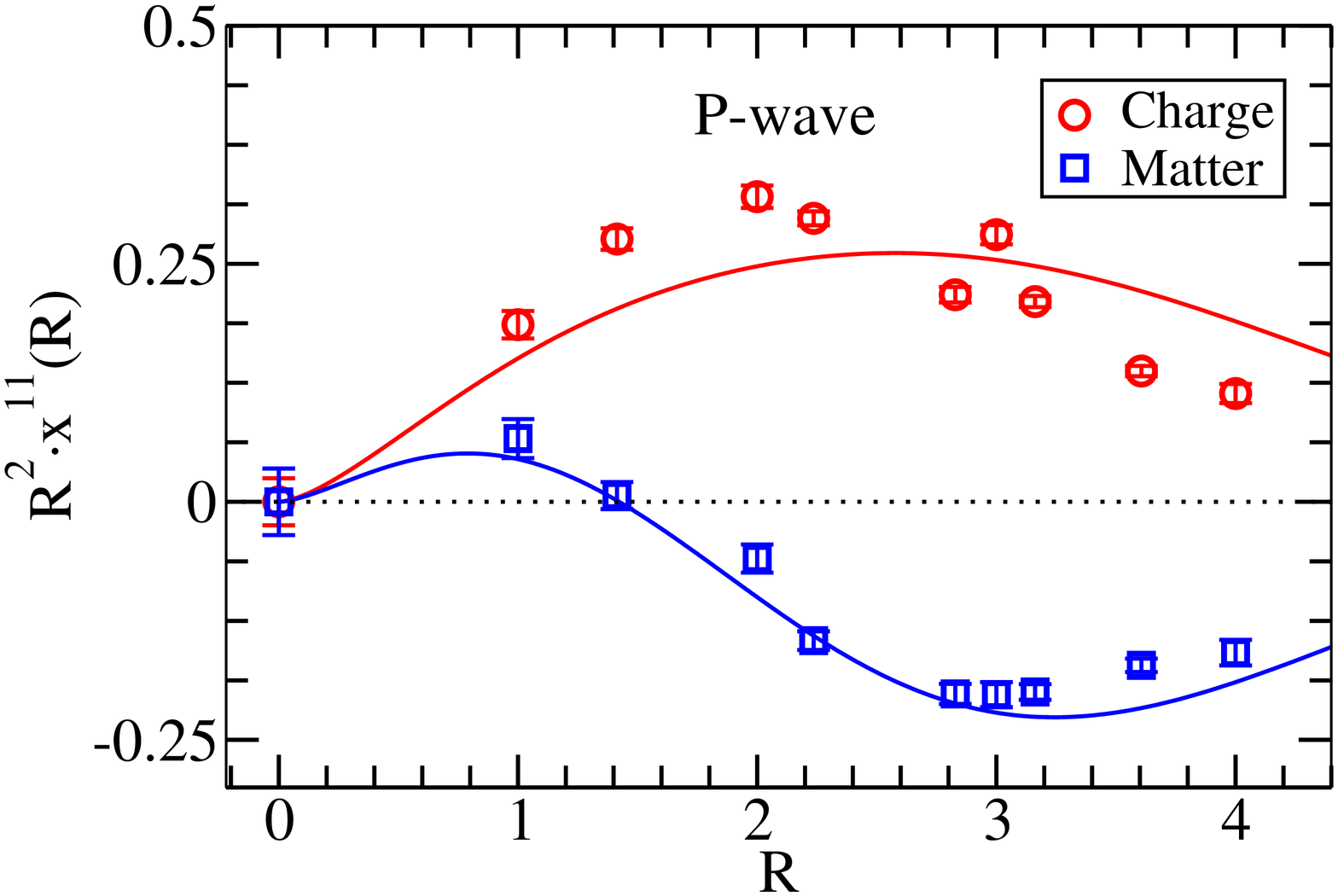}
\end{center}
\vspace{-9mm}
\label{fig_Dirac}
\end{figure}

The lattice measurements, \textit{i.e.} the energy spectrum and the radial
distributions, can be used to test potential models. There are several
advantages for model makers: First of all, since the heavy quark is infinitely
heavy, we have essentially a one-body problem. Secondly, on the lattice
we know what we put in --- one heavy quark and one light anti-quark ---
which makes the lattice system much more simple than the real world.
We can, for example, try to use the Dirac equation to interpret the S- and
P-wave distributions. Reasonable parameters can give a surprisingly
good qualitative fit to the distributions measured on the lattice ---
see Fig.~\ref{fig_Dirac}.

\section{Main conclusions}

\begin{itemize}
\item
There should be several narrow $B_s$ states,
\textit{e.g.} $0^+$ that lies below the $BK$ threshold.
\item
The spin-orbit splitting is very small.
\item
The radial distributions of S and P$-$ states can be qualitatively
understood by using a Dirac equation model.
\end{itemize}

\end{document}